\documentclass[12pt]{spieman}  
\usepackage{amsmath,amsfonts,amssymb}
\usepackage{graphicx}
\usepackage{setspace}
\usepackage{tocloft}

\usepackage{graphicx}
\usepackage{framed,multirow}
\usepackage[table]{xcolor}

\definecolor{beaublue}{rgb}{0.74, 0.83, 0.9}
\usepackage{tabulary}
\usepackage{multirow}
\usepackage{booktabs}
\usepackage{colortbl}
\usepackage{tabularx}
\usepackage{multirow}
\usepackage{threeparttable}
\usepackage{colortbl}
\usepackage{amssymb}
\usepackage{latexsym}

\usepackage{lineno,hyperref}
\modulolinenumbers[5]
\usepackage{hyperref}
\usepackage{multicol}
\usepackage{graphicx}
\usepackage{xcolor,cancel,mathrsfs,amscd}
\usepackage{subfig}
\usepackage{amsmath}
\usepackage{amssymb}
\usepackage{amsfonts}
\usepackage{multirow}
\usepackage{rotating}
\usepackage{subfig}
\usepackage{color}
\usepackage{float}
\usepackage{url}
\usepackage{breakcites}
\usepackage{array}
\title{Effect of Prior-based Losses on Segmentation Performance: A Benchmark}

\author[a,\textbf{*}]{Rosana El Jurdi}
\author[a]{Caroline Petitjean}
\author[b,c]{Veronika Cheplygina}
\author[a]{Paul Honeine} 
\author[d, e]{Fahed Abdallah}
\affil[a]{Normandie Univ, INSA Rouen, UNIROUEN, UNIHAVRE, LITIS, Rouen, France}
\affil[b]{Computer Science Department, IT University of Copenhagen, Denmark}
\affil[c]{Medical Image Analysis group, Eindhoven University of Technology, Eindhoven, The Netherlands}
\affil[d]{Luxembourg Institute of Socio-Economic Research (LISER), L-4366 Esch-sur-Alzette, Luxembourg}
\affil[e]{Universit\'e Libanaise, Hadath, Beyrouth, Liban }

\cftpagenumbersoff{figure}
\cftpagenumbersoff{table} 
\begin{document} 
\maketitle

\begin{abstract}
Today, deep convolutional neural networks (CNNs) have demonstrated state-of-the-art performance for medical image segmentation, on various imaging modalities and tasks. Despite early success, segmentation networks may still generate anatomically aberrant segmentations, with holes or inaccuracies near the object boundaries. To enforce anatomical plausibility, recent research studies have focused on incorporating expert knowledge also known as prior knowledge, such as object shapes or boundary, as constraints in the loss function. Prior integrated could be low-level referring to reformulated representations extracted from the ground-truth segmentations, or high-level representing external medical information such as the organ's shape or size. Over the past few years, prior-based losses exhibited a rising interest in the research field since they allow integration of expert knowledge while still being architecture-agnostic. However, given the diversity of prior-based losses on different medical imaging challenges and tasks, it has become hard to identify what loss works best for which dataset. In this paper, we establish a benchmark of recent prior-based losses for medical image segmentation. The main objective is to provide intuition onto  which losses to choose given a particular task or dataset, based on dataset characteristics and properties.
To this end, four low-level and high-level prior-based losses are selected. The considered losses are validated on 8 different datasets from a variety of medical image segmentation challenges including the Decathlon, the ISLES and the WMH challenge. The  proposed benchmark is conducted via  a unified segmentation network and learning environment. The considered prior-based losses are varied in conjunction with the Dice loss across the different datasets. Results show that whereas low level prior-based losses can guarantee an increase in performance over the Dice loss baseline regardless of the dataset characteristics, high-level prior-based losses can increase anatomical plausibility as per data characteristic.
\end{abstract}

\keywords{Prior-based loss functions, Prior-constrained CNNs, Medical Image Segmentation, Deep Learning}

{\noindent \footnotesize\textbf{*}Rosana EL JURDI,  \linkable{rosana.el-jurdi@univ-rouen.fr} }

\section{Introduction} \label{sect:intro} Medical image segmentation is the process of making per-pixel predictions in an image in order to identify organs or lesions from the background. Generally, medical images are largely versatile in nature, depending on the acquisition process and the type of object to be segmented. Imaging modalities include magnetic resonance imaging (MRI), computed tomography (CT), nuclear medicine functional imaging, ultrasound imaging, microscopy, to name a few. Hence, they vary in characteristics and nature and are broad with regards to the anatomical object of interest. As such, guaranteeing high performance for medical image segmentation can be considered very challenging when compared to other types of images or segmentation tasks. Regardless, segmentation in the medical domain is considered a key step in assisting early disease detection, diagnosis, monitoring treatment and follow up.

In the recent era, deep learning has registered a pivotal milestone in many fields including pattern recognition, object detection, natural language processing, with medical image segmentation being no exception to the rule. Convolutional neural networks (CNNs), a class of deep learning models, have been known to achieve considerable results due to their generalization ability. Since the segmentation process involves indicating not only what is present in an image but also where, medical image segmentation via CNNs considers a trade-off between contextual and spatial understanding. A pioneering approach for segmentation is the U-Net\cite{Ronneberger2015a} model, which is known for the ability to consider semantic and contextual information while achieving promising performance. U-Net has gained a high-level of success within image segmentation generally, and medical image segmentation particularly, due to its enhanced properties and powerful predictive notions. 

Despite undeniable success, segmentation networks for medical images, including U-Net and its variants, may still generate anatomically aberrant segmentations, with holes or inaccuracies near the object boundaries as demonstrated in papers\cite{Isensee2018, Bernard2018} . Thus, such models lack the anatomical plausibility and background that a medical expert has. Moreover, they often require large amounts of annotated training data, which is not easy to obtain in the medical domain. Un-annotated or partially labeled data are, rather, more easily available or less computationally expensive. 

To mitigate these limitations, recent research studies have focused on incorporating medical expert information, known as prior knowledge, as constraints within the deep learning framework. Prior knowledge can be information concerning the object shapes, size, topology, boundary or location, and has been known to be useful via variational approaches prior to the deep learning era. The exploitation of prior knowledge allows enforcing anatomical plausibility within segmentations provided by deep networks and can also overcome the need for fully labeled data \cite{Kervadec2019a,Kervadec2020} . 

Constraints via prior knowledge can be incorporated in CNNs either at the level of the network architecture \cite{ElJurdi2020, ElJurdi2020a, Trullo2017,Ghafoorian2016a,Ravishankar2017a, Khoreva2017a} or at the level of the loss function \cite{Kervadec2018, Kervadec2019d, Kervadec2020a, Shit2019, Caliva2019,Mirikharaji2018a, Arif2018a}. Whereas structural constraints are rather robust, loss constraints are more generic and can be plugged into any backbone network. Thus, prior-based loss functions offer a versatile way to include constraints at different scales, while maintaining interactions between regions as well as the computational efficiency of the backbone network. 

Integrated prior can be low-level, which  resembles reformulated ground-truth representation and is extracted from the ground-truth segmentations. For example, distance maps \cite{Caliva2019, Kervadec2019a, Mosinska2018} and Laplacian filters \cite{Arif2018a}. Prior could also be high-level representing actual external medical information such as the shape of the organ, compactness  or size, and are optimized directly based on ground-truth prior tags \cite{Kervadec2018, Dolz2017a, Mirikharaji2018a}. 


Over the past few years, prior-based losses, whether low-level or high-level, present a rising trend in today's research in semantic image segmentation, particularly in the medical field. Given the diversity of prior-based losses on different medical imaging challenges and tasks, it has become hard to identify what loss works best for which dataset. For this reason, we establish in this paper, a benchmark of recent prior-based losses for medical image segmentation. Our main objective is to provide intuition onto  which losses to choose given a particular task or dataset, based on dataset characteristics and properties. 

In the literature, benchmarks most relevant to ours is the one proposed in paper \cite{Ma2020} and paper \cite{MA2021}. In  paper \cite{MA2021} , a benchmark of 20 losses is conducted with a thorough comparison on 4 main segmentation tasks: Liver, Liver Tumor, Pancreas and Multi-Abdominal Organ Segmentation. However, the authors do not address prior-based losses. Instead,  they consider regular fitting losses like Dice, Cross entropy and their variants. Their benchmark is limited to only 4 datasets. The benchmark proposed in paper\cite{Ma2020} targets some low-level prior-losses. However, this benchmark is limited to the scope of losses based on distance maps, such as the boundary loss \cite{Kervadec2019a} or the Hausdorff loss \cite{Karimi2019b} , and do not compare relative to high-level prior losses. Benchmark \cite{Ma2020} also demonstrate results on some structural constraints (i.e., regarding the architecture), that do not lie within the scope of our work. In addition, the benchmark is limited to two datasets: an organ segmentation task of the left atrial structure within MRI images and a liver tumor segmentation task within CT scans. In this work, we target specifically prior-based losses, both high-level and low-level, on 8 datasets of different tasks and modalities. Hence up to our knowledge, there is no benchmark that aims to compare prior-based losses on a number of datasets in order to quantify common trends and limitations.

The main objective of the proposed benchmark is to study the performance of prior-based losses, on a variety of datasets, tasks and modalities. In this way, we provide the readers with intuition onto which losses to choose given a particular task of interest. Prior-based losses are quite interesting because they allow integration of expert knowledge while still being architecture-agnostic, that is to say, they can be plugged into any backbone network. As a result, we are able to unify the segmentation network given the same learning environment, while varying the prior-based losses accordingly. We note that each of the considered losses has been proposed in their respective papers, in order to carry on a particular task. We believe that aside from the initial motive that the considered losses were designed for, additional significance may be drawn on other segmentation tasks and dataset characteristics. For this reason, we validate the chosen prior-based losses on 8 different datasets from a variety of medical image segmentation challenges including the Decathlon\footnote{\href{http://medicaldecathlon.com/}{http://medicaldecathlon.com/}} , the ISLES \footnote{\href{http://www.isles-challenge.org/}{http://www.isles-challenge.org/}} and the WMH\footnote{\href{https://wmh.isi.uu.nl/}{https://wmh.isi.uu.nl/}}  challenge. The main contributions of this paper are summarized as follows: 
\begin{itemize}
	\item We present a benchmark of architecture-agnostic prior-based losses for medical image segmentation.
	\item We attempt to shed light on the underlying relationship between the prior-based losses and some dataset characteristics.
	
\end{itemize}

The rest of the paper is organized as follows. Section \ref{benchmark-lossfunctions} presents the selected loss functions for the proposed benchmark and elaborate on why the proposed losses were chosen. Section \ref{benchmark-expSetting} illustrates the experimental setting adopted in order to evaluate the considered prior-based losses on the different datasets.  In section \ref{Benchmark-Datasets}, we describe the datasets considered and the meta-features extracted to compare the loss performances. Finally, section \ref{benchmark-results} demonstrates the results and analyzes the loss performances relative to segmentation tasks and dataset characteristics.

\section{Selected Loss Functions} \label{benchmark-lossfunctions}
In this section, we present the chosen prior-based losses for our proposed benchmark.
Prior-based losses can be high-level, when the type of prior considered is based on external  knowledge (e.g. shape), or low-level, that integrate ground-truth map transformations such as distance or contour maps, in order to reveal geometrical and location properties as demonstrated in survey \cite{ElJurdi2021} . The proposed benchmark mainly focuses on 4 recent prior-based losses that have raised interest within the field of medical image segmentation, 2 are low-level, and the other 2 are high-level.

\subsection{Low-level prior-based losses}
Possible low-level prior can be based on distance map as demonstrated in papers \cite{Kervadec2019a, Karimi2019b} . In this context, two major contributions are the boundary loss \cite{Kervadec2019a} and the Hausdorff loss \cite{Karimi2019b} . \\

The \textbf{Boundary loss} $\mathcal{L}_{Boundary}$ is an approximation of the distance between the real and the estimated boundaries. Based on graph theories\cite{Boykov2001a} , an equivalent term that finetunes the probability distribution via ground-truth distance maps is derived in paper \cite{Kervadec2019a} and is defined as:
\begin{equation}
	\small
	\mathcal{L}_{Boundary} = \sum_{p \in \Omega} \phi_g(p). \hat{y}_p \qquad \quad 
	\text{with } \phi_g(p) = 
	\begin{cases}
		-D_G(y_p) &  \text{for } p \text{ inside the target region } \\
		D_G(y_p) & \text{else},
	\end{cases}
	\label{kerva_boundary}
\end{equation}
where $D_G(p)$ denotes the distance of pixel $p$ to the closest contour ($G$) point,  $\widehat{y}_p$ being the predicted value at pixel $p$, and $\Omega$ the image spatial domain.\\

The \textbf{Hausdorff loss} $\mathcal{L}_{HD}$ \cite{Karimi2019b} conducts a direct point-by-point optimization of the predicted and ground-truth contours arriving to the following loss term:
\begin{equation}
	\mathcal{L}_{HD} = \frac{1}{|\Omega|} \sum_{p \in \Omega}(y_p - \hat{y}_p)^2 \Big (D_G(y_p)^{2} + D_G(\hat{y}_p)^{2} \Big  ).
	\label{hauss}
\end{equation}

The boundary loss has been initially designed in order to segment lesions within the brain, with the WMH and the ISLES datasets, whereas the Hausdorff loss has been tested on 4 different single-organ segmentation tasks, including the prostate, liver and pancreas from the Decathlon and PROMISE challenges. However, these losses were not evaluated in multi-organ segmentation. Since both losses lie in the same spectrum of low-level prior-based losses, and rely on the distance map, it may be interesting to investigate their performance on  the same datasets in order to pinpoint common behaviors. Moreover, we aim to also extend the scope of these losses to the multi-organ case. 



\subsection{High-level prior-based losses}

Regarding high-level prior losses, we analyze the performance of the clDice loss \cite{Shit2019} and the size loss \cite{Kervadec2018} . \\

The \textbf{Size loss} \cite{Kervadec2018} estimates the organ size from a soft probability map and constrains it, based on higher and lower threshold value of the organ size, according to the following: 

\begin{equation}
	A(\widehat{y}) = \sum\limits_{p \in \Omega}\widehat{y}_p,
	\label{area}
\end{equation}

\begin{equation}
	\mathcal{L}_{size} = 
	\begin{cases}
		\big(A(\widehat{y}) - a \big)^2 & \text{if } A(\widehat{y}) \leq a,\\
		\big(A(\widehat{y}) - b \big)^2 & \text{if } A(\widehat{y}) \geq b,\\
		0 & \text{otherwise},
	\end{cases}
	\label{eq:size-loss}
\end{equation}

\noindent
where $a$ and $b$ are respectively the upper and lower permissible bounds that the size of the considered object can attain.  The size loss was originally designed for weakly supervised learning, to guide the network through the training despite the  lack of full label maps. We are particularly interested in studying the effect of the size loss on small structures that are known to be more difficult to segment.\\


\smallskip
The \textbf{clDice loss} \cite{Shit2019} , also called skeleton loss, exploits skeletonization maps that are compact representations of images and objects that preserve topological properties. The objective of this loss is to constrain the skeleton of the predicted map to match the skeleton of the ground-truth map. This prior was used in the segmentation of vessels and neurons in both 2D and 3D.
Let $s$ and $\widehat{s}$ be the ground-truth and the predicted skeleton respectively, of size $|\Omega|$. The sensitivity (or recall) between the predicted segmentation and ground-truth  skeleton is introduced as $T_{sens}(s, \widehat{y}) =  {|s \cap \widehat{y}|}/{|s|}$.
Likewise, the precision between the ground-truth mask $y$ and the predicted skeleton $\widehat{s}$ is defined as: $T_{prec}(\widehat{s}, y) = {|\widehat{s} \cap y|}/{|\widehat{s}|}$. The clDice is defined as the F1-score between precision $T_{prec}$ and sensitivity $T_{sens}$ as follows:  
\begin{equation}
	\mathcal{L}_{clDice} = 2 \frac{T_{prec}(\widehat{s}, y) \, T_{sens}(s, \widehat{y})}{T_{prec}(\widehat{s}, y) + T_{sens}(s, \widehat{y}) }.
	\label{skeleton-loss}
\end{equation}

The clDice was originally designed to segment vessels; however, due to the nature of the skeletonization feature that they target, we believe that they may be good at distinguishing between different structures lying in close proximity to each other, such as when the organs are made of multiple instances.

\begin{figure*}[t!]
	\subfloat[][White Matter Hyper-intensity dataset (WMH) from the MICCAI challenge \cite{Kuijf2019}]{
		
		\begin{minipage}[c]{0.47\textwidth}
			\centering
			\includegraphics[width=7cm,height=3.5cm]{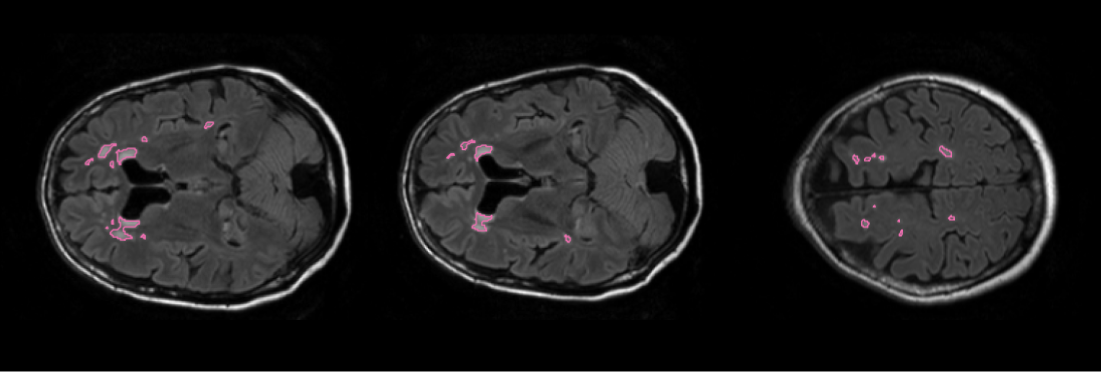} \label{fig:WMH} \\
		\end{minipage}
	}
	\hfill 	
	\subfloat[][Ischemic Stroke Lesion Segmentation (ISLES) dataset from the MICCAI 2015 challenge \cite{Maier2017}]{
		\begin{minipage}[c]{0.47\textwidth}
			\centering
			\includegraphics[width=7cm,height=3.5cm]{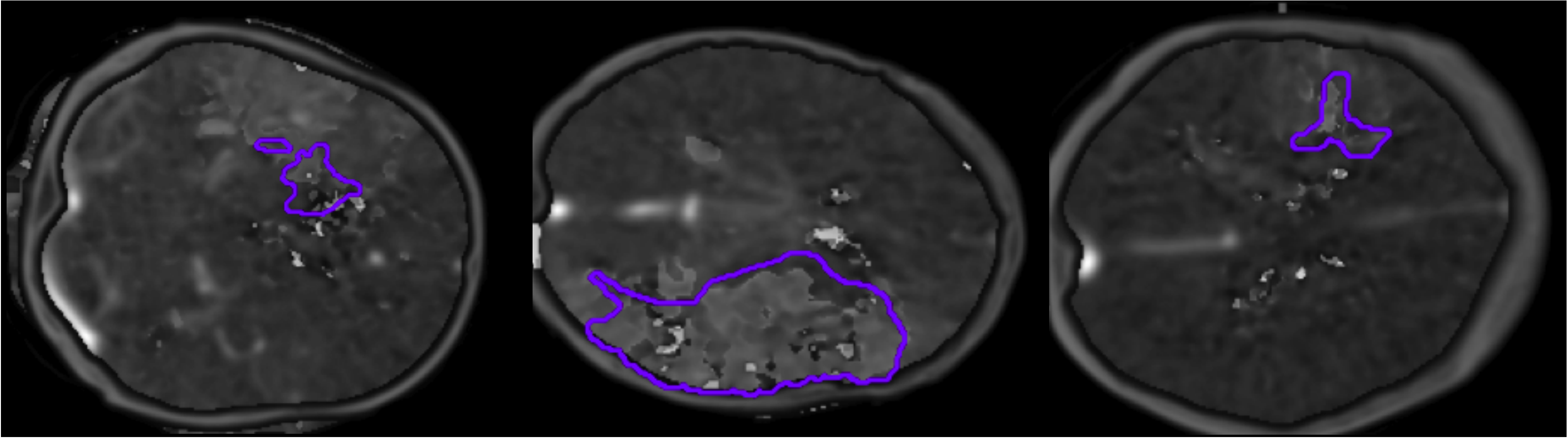} \label{fig:ISLES}  
	\end{minipage}}
	\caption{Brain lesion segmentation task}
	\label{fig:Brain-Lesion}
	
\end{figure*}



\begin{table*}[t!]
	\caption{Dataset Description: \# of patients: patient split is 80 \% / 20 \% on the original dataset; Organ Size: \% of pixels occupied by the organ w.r.t. the entire image;  \# of CC: number of connected components;}
	\smallskip
	\begin{center}
		\begin{tabular}{|lc|cc|cc|c|c|c|}\cline{3-9}
			\multicolumn{2}{c|}{}& \multicolumn{2}{c}{\# of patients} & \multicolumn{2}{c}{Organ Size} & \# of classes & \# of modality &  \# of \\
			\multicolumn{2}{c|}{}& Train & Test &  mean & std & &  & CC \\\hline
			WMH & & 48 & 12 & 0.33 & 0.56 &1 & 2 &  0 $ \sim $ 26  \\ \hline
			Isles & & 74 & 20 & 2.11 & 1.91 & 1 & 5&   0 $ \sim $ 3  \\ \hline
			Atrium & &16 & 4 & 0.69 & 0.43 & 1 & 1 &   0 $ \sim $ 4  \\  \hline
			Colon & & 100 & 38 & 0.6 & 0.59 & 1 & 1 &  0 $ \sim $ 3  \\ \hline
			Spleen  & & 32 & 9  & 1.57 & 1.03  &1 & 1 &  0 $ \sim $ 1  \\ \hline
			
			\multirow{2}{*}{Hippo.}  & \textbf{\textit{H1}}  & \multirow{2}{*}{206} & \multirow{2}{*}{54} & 4.08 & 3.87 & \multirow{2}{*}{2} & \multirow{2}{*}{1}  &  
			0 $ \sim $ 4  \\ 
			& \textbf{\textit{H2}}  & & &  3.53 & 2.53 & & &  0 $\sim$ 3  \\ \hline
			
			\multirow{2}{*}{Prost.} & \textbf{\textit{CG}} & \multirow{2}{*}{26} & \multirow{2}{*}{6} & 0.9 & 0.89 & \multirow{2}{*}{2} & \multirow{2}{*}{2}  &
			$0 \sim 26  $  \\ 
			& \textbf{\textit{PZ}}  & & &  3.1 & 2.98 & & &  0 $ \sim  $ 1\\ \hline	
			
			\multirow{3}{*}{ACDC} &   \textbf{\textit{RVC}} \textbf{}&\multirow{3}{*}{99} & \multirow{3}{*}{24}  & 1.29 & 1.03 & \multirow{3}{*}{3} & \multirow{3}{*}{1} &  \multirow{3}{*}{0 $\sim$ 1}  \\
			& \textbf{\textit{MYO}} & & & 1.38 & 0.69& & & \\
			& \textbf{\textit{LVC}}  & & & 1.28 & 0.84 & & &  \\ \hline
			
		\end{tabular}
	\end{center}
	\label{tab:metadata}
\end{table*}

\section{Datasets and Tasks} \label{Benchmark-Datasets}
In this section, we present a brief description of the datasets under consideration. The datasets were chosen to cover different tasks, modalities and characteristics. Each dataset encompasses a particular set of challenges the segmentation network must consider while training.
A summary of the meta-dataset characteristics is presented in \tablename~\ref{tab:metadata}.


\begin{figure*}[t!]
	\subfloat[][Spleen Dataset from Decathlon challenge  showing  organ or large size variability and convexity issues at boundary level.]{
		
		\begin{minipage}[c]{0.3\textwidth}
			\centering
			\includegraphics[width=5cm, height=3cm]{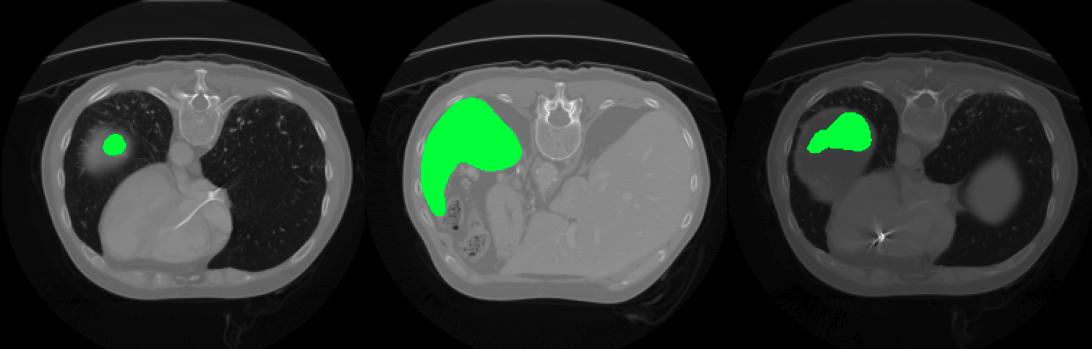} \label{fig:topo-examples}
		\end{minipage}
	}
	\hspace{0.2cm}
	\subfloat[][MR images from the Atrium dataset with manual segmentation and bounding boxes segmentation overlaid on the left atrium.]{
		\begin{minipage}[c]{0.3\textwidth}
			\centering
			\includegraphics[width=5cm, height=3cm]{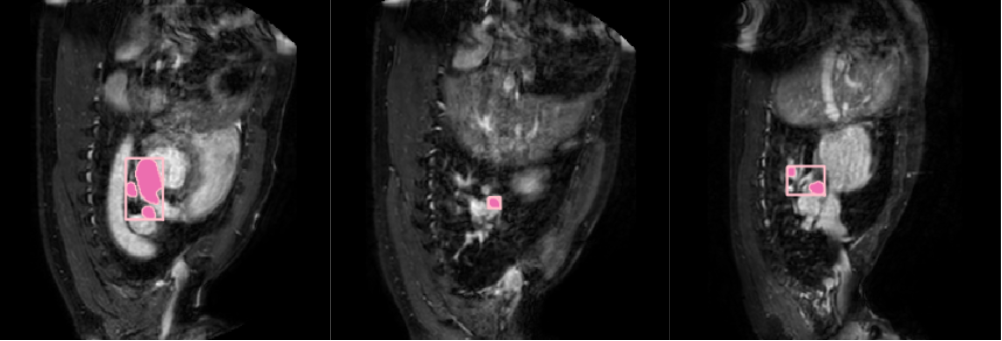} \label{fig:interregions-examples} 
	\end{minipage}} \hspace{0.2cm}
	\subfloat[][Colon Dataset  showing  Colon Cancer Primaries of different curvatures and sizes.]{
		\begin{minipage}[c]{0.3\textwidth}
			\begin{center}
				\includegraphics[width=5cm, height=3cm]{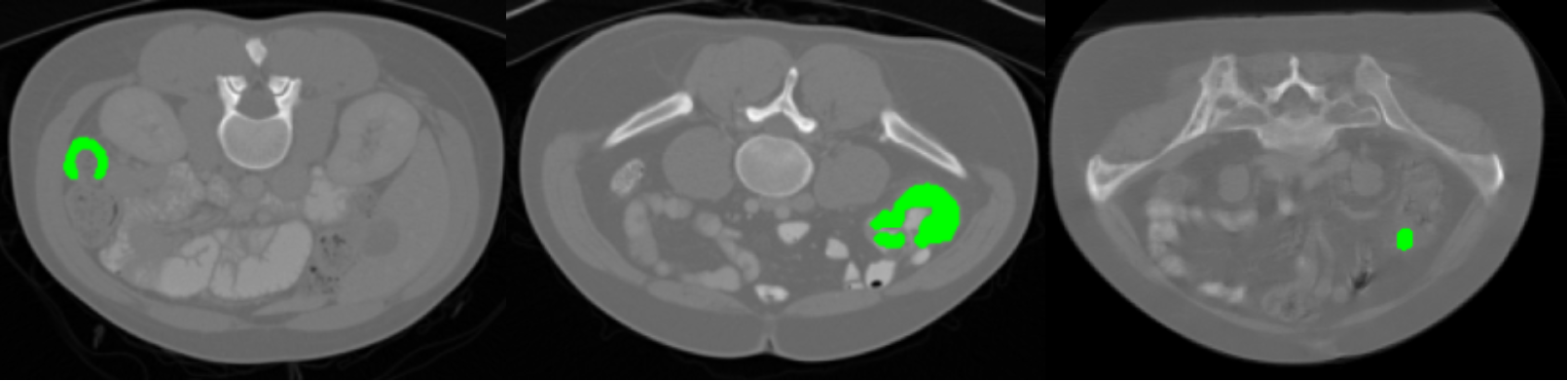} \label{fig:interregions-examples} 
			\end{center}
			
	\end{minipage}}
	
	\caption{Single-organ segmentation tasks from the Decathlon Challenge \cite{Simpson2019a}. }
	\label{fig:Single-Organ-Seg}
	
\end{figure*}
\subsection{Brain Lesion Segmentation}

To investigate the significance of prior-based losses on Brain lesion segmentation tasks, we mainly focus on the segmentation of white matter hyperintensities (WMH) dataset and the ischemic stroke lesion segmentation dataset (ISLES). Both datasets are multi-modal with anatomical objects that are characterized by being sparse and composed of multi-instances (See \figurename~\ref{fig:Brain-Lesion}).

\subsection{Single Organ Segmentation}

Organs can generally be single-connected of only 1 structure, or multi-connected composed of multi structures that are close to each other. To investigate the segmentation performance of prior-based losses on single-organ segmentation tasks where the organ considered is characterized with multi-connected structures, we targeted the segmentation of the atrium and Colon from the Decathlon Challenge. Alternatively, we target the spleen to investigate the performance of prior-based losses relative to single-label single-connected organs. The spleen and colon are characterized with a largely varying size and mild convexity issues at boundary levels. On the other hand, the atrium is a multi-instance anatomical object with up to 4 elements of varying sizes and lying in close proximity to each other (See \figurename~\ref{fig:Single-Organ-Seg}).

\begin{figure*}[t!]
	\subfloat[][Prostate Dataset  showing  central gland (pink) and peripheral zone (green) \cite{Simpson2019a}.]{
		
		\begin{minipage}[c]{0.3\textwidth}
			\centering
			\includegraphics[width=5cm, height=3cm]{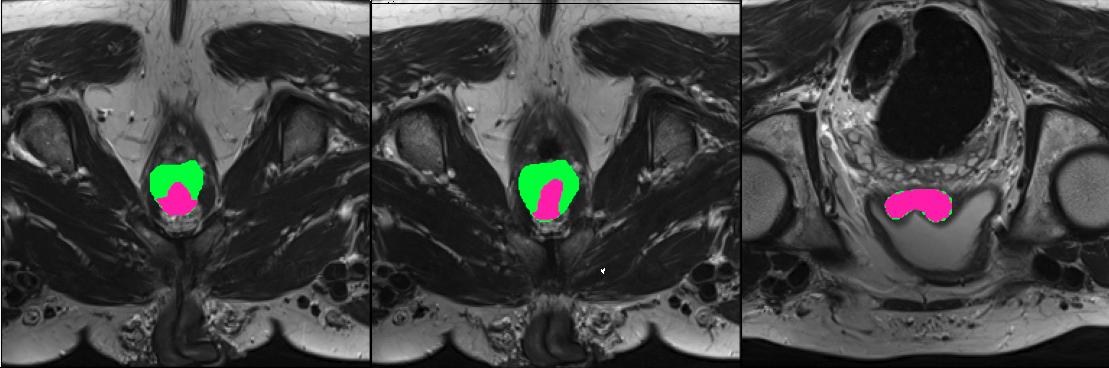} \label{fig:topo-examples}
		\end{minipage}
	}
	\hspace{0.2cm}
	\subfloat[][ACDC Dataset with right ventricle structure is  in blue, left ventricle structure is yellow and myocardium structure in green \cite{Bernard2018}]{
		\begin{minipage}[c]{0.3\textwidth}
			\centering
			\includegraphics[width=5cm, height=3cm]{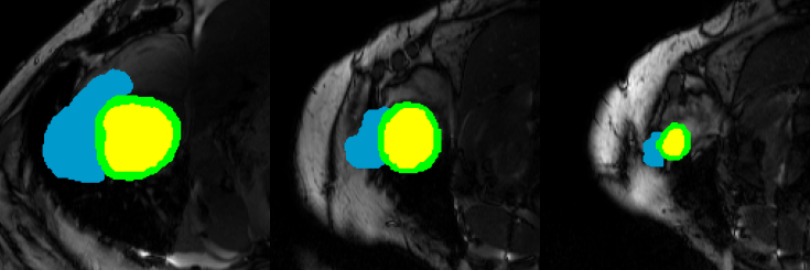} \label{fig:interregions-examples} 
	\end{minipage}} \hspace{0.2cm}
	\subfloat[][Hippocampus Dataset from Decathlon challenge where two brain tissues lying in close proximity of each other are to be segmented given low resolution images \cite{Simpson2019a}.]{
		\begin{minipage}[c]{0.3\textwidth}
			\begin{center}
				\includegraphics[width=5cm, height=3cm]{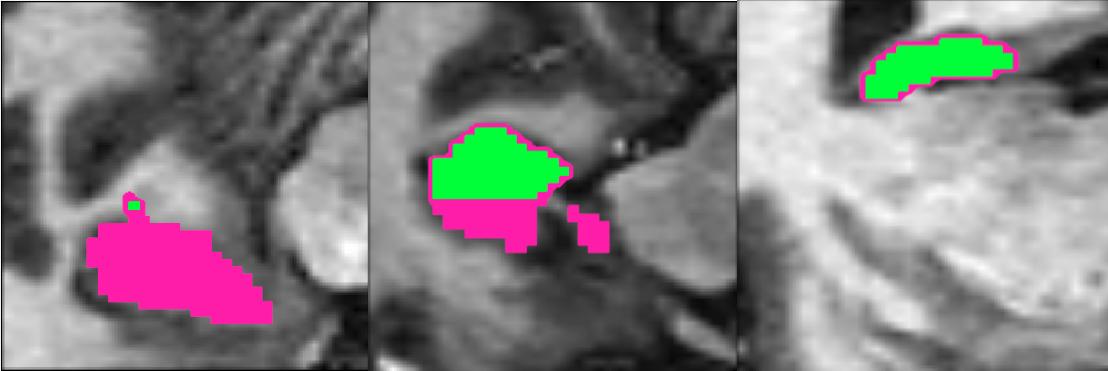} \label{fig:interregions-examples} 
			\end{center}
			
	\end{minipage}}
	
	\caption{Multi-Organ Segmentation Tasks.}
	\label{targetobject}
	
\end{figure*}

\subsection{Multi-Organ Segmentation}
For multi-organ segmentation, we have targeted the Prostate  (Prostate central gland and peripheral zone) and Hippocampus (tissues H1 \& H2) datasets from the Decathlon Challenge and the ACDC dataset (Three Cardiac Structures).

\subsection{Meta-dataset Features}

In order to reveal the underlying relationship between loss performance and dataset characteristics, we propose a set of meta-features that describe the datasets. This includes the size of the anatomical object taken as the percentage of occupation from the entire image, the number of connected components, which means how many instances an anatomical object is constituted of, and the number of classes, i.e., whether the segmentation task is single or multi-label.

\section{Experimental Setting}  \label{benchmark-expSetting}

We deploy a unified U-Net based framework \cite{Kervadec2019a,Kervadec2018} and modify the loss function accordingly. Training is done using a batch size of 8 and a learning rate of $10^{-3}$.  The learning rate is halved if the validation performance does not improve during 20 epochs. The U-Net model is trained via each prior-based loss in conjunction with the Dice loss weighted by a parameter $\lambda $ according to the following equation:

\begin{equation}
	\mathcal{L} =(1-\lambda) \mathcal{L}_{Dice} + \lambda \mathcal{L}_{prior}
	\label{eq:dynamic-training}
\end{equation} 
The parameter  $\lambda $ is fine-tuned via the dynamic training strategy \cite{Kervadec2019a} such that its value was initially set to 0.01 and increased by 0.01 per epoch for 200 epochs. Our code is publically available on GitHub \footnote{\href{https://github.com/rosanajurdi/Prior-based-Losses-for-Medical-Image-Segmentation}{https://github.com/rosanajurdi/Prior-based-Losses-for-Medical-Image-Segmentation}} 



For pre-processing, we have resized the images to 256 $\times$ 256 pixels and normalized them to the range [0, 1].
For multi-modal datasets, we have concatenated the channels at the level of the input. Each dataset was split into train and validation based on an 80 \% / 20 \% partition respectively, as shown in \tablename~\ref{tab:metadata}, and validated via three Monte-Carlo simulations \cite{Arlot2010}.



\begin{table*}[t!]
	\caption{Average Dice scores $\pm$ standard deviation. 
		Blue (resp. pink) background represents Dice Accuracy superior (resp. inferior) to the corresponding Dice baseline. The bold result is the best Dice score (i.e., the greatest) obtained on the dataset.
	}
	\smallskip
	\begin{center}
		\scalebox{0.9}{
			\begin{tabular}{lc|c|cccccc}
				Dataset  & & $\mathcal{L}_{Dice}$ & $\mathcal{L}_{Dice}$+$ \mathcal{L}_{Boundary}$& $\mathcal{L}_{Dice}$+$\mathcal{L}_{HD}$  & $\mathcal{L}_{Dice}$+ $\mathcal{L}_{Size}$ &$\mathcal{L}_{Dice}$ + $\mathcal{L}_{clDice}$\\ \hline \hline
				WMH & &  74.64 $\pm$ 1.34  & \cellcolor{beaublue}{ 77.29 $\pm$  0.75}  &   \cellcolor{beaublue}{\textbf{78.77} $\pm$\textbf{ 0.70 }}   & \cellcolor{beaublue}{\textbf{78.06} $\pm$ \textbf{1.61}}  &  \cellcolor{red!10}{66.97 $\pm$ 11.48}   \\ 
				
				Isles & &  53.41 $\pm$ 4.61  &  \cellcolor{beaublue}{ 62.93 $\pm$  2.24 } & \cellcolor{beaublue}{\textbf{63.53} $\pm$ \textbf{1.66}}  & \cellcolor{red!10}{46.86 $\pm$ 7.74}  & \cellcolor{beaublue}{62.53 $\pm$ 5.22}    \\
				
				
				Atrium  &  & 83.67 $\pm$ 3.66  & \cellcolor{beaublue}{ 82.80 $\pm$  3.68} & \cellcolor{beaublue}{\textbf{84.57} $\pm$ \textbf{1.86}}
				& 
				\cellcolor{beaublue}{\textbf{84.59} $\pm$ \textbf{2.62}}  & \cellcolor{beaublue}{83.85 $\pm$ 2.56} \\ 
				
				Colon &  & 84.82 $\pm$ 1.71  &  \cellcolor{beaublue}{\textbf{88.71 $\pm$  0.48}} & \cellcolor{beaublue}{\textbf{88.30} $\pm$ \textbf{0.78} }  & \cellcolor{beaublue}{\textbf{88.71} $\pm$ \textbf{0.48}}  &  \cellcolor{beaublue}{84.52 $\pm$ 2.64}  \\  
				
				Spleen &  & 76.80 $\pm$ 7.59  &  \cellcolor{beaublue}{80.38 $\pm$  5.46}  &  \cellcolor{beaublue}{\textbf{91.79 $\pm$ 2.67} }    & 
				\cellcolor{beaublue}{86.44 $\pm$ 15.86 } &  \cellcolor{beaublue}{87.15 $\pm$ 13.61}    \\ 
				
				
				\hline
				\multirow{2}{*}{Hippocampus}& \textit{H1} &  49.38 $\pm$ 0.33  &  \cellcolor{beaublue}{65.20 $\pm$  0.31}  & 
				\cellcolor{beaublue}{\textbf{68.54 $\pm$ 1.46} } & 
				\cellcolor{beaublue}{66.24 $\pm$ 0.33}  &  \cellcolor{beaublue}{\textbf{68.39 $\pm$ 2.60}}  \\
				& \textit{H2} &  71.70 $\pm$ 1.30  & \cellcolor{beaublue}{ 81.33 $\pm$  0.74}  & 
				\cellcolor{beaublue}{\textbf{82.12} $\pm$ \textbf{0.44}}  & 
				\cellcolor{beaublue}{81.84 $\pm$ 0.63 }  &  \cellcolor{beaublue}{\textbf{82.82 $\pm$ 1.22} }  \\ 
				
				\hline

				\multirow{2}{*}{Prostate}	&  \textit{CG}& \textbf{45.17 }$\pm$ \textbf{6.41 } &  \cellcolor{red!10}{44.89 $\pm$  7.09}  &  \cellcolor{red!10}{44.15 $\pm$ 5.61}   &  \cellcolor{red!10}{34.12 $\pm$ 7.49}  &  \cellcolor{red!10}{42.45 $\pm$ 7.03}   \\
				
				& \textit{PZ} &  65.13 $\pm$ 11.57  & \cellcolor{beaublue}{ \textbf{68.99} $\pm$  \textbf{9.94 }} & 
				\cellcolor{red!10}{ 64.38 $\pm$ 9.33   } & \cellcolor{red!10}{29.61 $\pm$ 12.07}  &  \cellcolor{red!10}{61.57 $\pm$  11.44}   \\ 
				\hline
				\multirow{3}{*}{ACDC}& \textit{RVC} & 80.79 $\pm$ 0.95 & \cellcolor{beaublue}{81.04 $\pm$ 0.87}  &  \cellcolor{red!10}{80.54 $\pm$ 1.30} &  \cellcolor{red!10}{41.02 $\pm$ 38.39 }& \cellcolor{beaublue}{\textbf{83.83 $\pm$ 1.39}} \\
				
				& \textit{MYO}&  \textbf{83.92 $\pm$ 0.13} & \cellcolor{beaublue}{\textbf{84.16 $\pm$ 0.83}} & \cellcolor{red!10}{83.91 $\pm$ 0.85}  & \cellcolor{red!10}{83.41 $\pm$ 0.72} & \cellcolor{red!10}{83.24 $\pm$ 0.66} \\
				& \textit{LVC}  & \textbf{90.26 $\pm$ 0.13} & \cellcolor{red!10}{89.53 $\pm$ 0.74} & \cellcolor{red!10}{88.98 $\pm$ 0.90 }  & \cellcolor{red!10}{ 89.74 $\pm$ 0.71}  & \cellcolor{red!10}{89.56 $\pm$ 1.10} \\ \hline
		\end{tabular}}
	\end{center}
	\label{tab:BenchmarkResults-DSC}
\end{table*}
\begin{table*}[t!]
	\caption{Average  Hausdorff Distances $\pm$ standard deviation.	Blue (resp. pink) background represent HD inferior (resp. superior) to the corresponding Dice baseline. The bold result is the best (i.e. the smallest) Hausdorff Distance obtained on the dataset.}
	\smallskip
	\begin{center}
		\scalebox{0.95}{
			\begin{tabular}{lc|c|cccccc}
				Data-Set  & & $\mathcal{L}_{Dice}$ & $\mathcal{L}_{Dice}$+$ \mathcal{L}_{Boundary}$& $\mathcal{L}_{Dice}$+$\mathcal{L}_{HD}$ & $\mathcal{L}_{Dice}$+ $\mathcal{L}_{Size}$ &$\mathcal{L}_{Dice}$ + $\mathcal{L}_{clDice}$  \\ 
				
				\hline
				WMH  & &  0.98 $\pm$ 0.13  &   \cellcolor{beaublue}{0.94 $\pm$  0.17} & \cellcolor{beaublue}{\textbf{0.93} $\pm$ \textbf{0.16}}  & \cellcolor{beaublue}{0.94 $\pm$ 0.18}  &  \cellcolor{red!10}{1.16 $\pm$ 0.38}  \\ 
				
				Isles & & 3.75 $\pm$ 0.35  &  \cellcolor{beaublue}{\textbf{3.05} $\pm$  \textbf{0.22}} & \cellcolor{beaublue}{3.07 $\pm$ 0.18}    &  \cellcolor{beaublue}{3.45 $\pm$ 0.79} & \cellcolor{beaublue}{3.29 $\pm$ 0.62}   \\  
				Atrium & & 1.62 $\pm$, 0.16  &  \cellcolor{red!10}{1.64 $\pm$  0.16}  & \cellcolor{red!10}{1.67 $\pm$ 0.13}   & 
				\cellcolor{beaublue}{\textbf{1.59} $\pm$ \textbf{0.17}} & \cellcolor{red!10}{ 1.64 $\pm$ 0.16}   \\ 
				
				Colon & & 0.58 $\pm$ 0.04  &  \cellcolor{beaublue}{\textbf{0.50} $\pm$  \textbf{0.02}} & \cellcolor{beaublue}{0.51 $\pm$ 0.03}   & \cellcolor{beaublue}{\textbf{0.50} $\pm$ \textbf{0.02}}  & \cellcolor{red!10}{ 0.58 $\pm$ 0.07 } \\
				Spleen & & $ 1.33 \pm, 0.28 $ & \cellcolor{red!10}{$ 1.34 \pm  0.21 $} & \cellcolor{beaublue}{\textbf{0.92} $\pm$ \textbf{0.15 }}    & \cellcolor{beaublue}{$
					1.27 \pm 0.57 $} & \cellcolor{beaublue}{ 1.07 $\pm$ 0.53  }\\ 
				\hline
				\multirow{2}{*}{Hippocampus} & \textit{H1} & 2.31 $\pm$ 0.05  & \cellcolor{beaublue}{ 1.99 $\pm$  0.01} &
				\cellcolor{beaublue}{1.98 $\pm$ 0.03}  &
				\cellcolor{beaublue}{\textbf{1.97} $\pm$ \textbf{0.02}}  &  \cellcolor{beaublue}{1.99 $\pm$ 0.04}   \\
				& \textit{H2} &  3.82 $\pm$ 0.14  & \cellcolor{beaublue}{ 3.09 $\pm$  0.01 } &
				\cellcolor{beaublue}{\textbf{2.97} $\pm$ \textbf{0.05}}  &
				\cellcolor{beaublue}{3.07 $\pm$ 0.01}  & \cellcolor{beaublue}{ 3.20 $\pm$ 0.18}  \\ 
				\hline
				\multirow{2}{*}{Prostate} & \textit{CG} &  2.80 $\pm$ 0.34 & \cellcolor{beaublue}{\textbf{2.77} $\pm$  \textbf{0.43}} & \cellcolor{red!10}{2.88 $\pm$ 0.27 } &
				\cellcolor{red!10}{3.10 $\pm$ 0.26} & \cellcolor{red!10}{3.48 $\pm$ 0.66}   \\ 
				& \textit{PZ} &  3.24 $\pm$ 0.35 &\cellcolor{beaublue}{\textbf{2.94} $\pm$  \textbf{0.27}} & \cellcolor{beaublue}{3.17 $\pm$ 0.47} & \cellcolor{red!10}{4.41 $\pm$ 0.90}  & \cellcolor{red!10}{3.45 $\pm$ 0.58 } \\

				\hline

				\multirow{3}{*}{ACDC} & \textit{RVC}  & 2.44 $\pm$ 0.04 & \cellcolor{beaublue}{2.41 $\pm$ 0.05} & \cellcolor{beaublue}{\textbf{2.33} $\pm$ \textbf{0.04}}   & \cellcolor{red!10}{3.88 $\pm$ 1.44}  & \cellcolor{beaublue}{2.34 $\pm$ 0.08}\\
				
				& \textit{MYO}&  2.60 $\pm$ 0.01  & \cellcolor{beaublue}{\textbf{2.57} $\pm$ \textbf{0.01}}  & \cellcolor{red!10}{2.65 $\pm$ 0.01 } & \cellcolor{red!10}{2.62 $\pm$ 0.00}  & \cellcolor{red!10}{2.71 $\pm$ 0.04} \\ 
				&\textit{LVC} & 1.95 $\pm$ 0.02 & \cellcolor{red!10}{1.95 $\pm$ 0.02} & \cellcolor{red!10}{1.98 $\pm$ 0.01}   & \cellcolor{red!10}{1.94 $\pm$ 0.01 } &\cellcolor{red!10}{ 1.98 $\pm$ 0.04}  \\ 
				\hline
		\end{tabular}}
	\end{center}
	\label{tab:Results-HD}
\end{table*}

\begin{table*}[h!]
	\caption{ Mean Absolute Error (MAE) on the number of connected components (CC) of the ground truth vs the number of CC of the predicted segmentation map. 	Blue (resp. pink) background represents an MAE inferior (resp. superior) to the corresponding Dice baseline. }
	\smallskip
	\begin{center}
		\scalebox{0.99}{
			\begin{tabular}{l|c|cccccc}
				Data-Set  & $\mathcal{L}_{Dice}$ & $\mathcal{L}_{Dice}$+$ \mathcal{L}_{Boundary}$& $\mathcal{L}_{Dice}$+$\mathcal{L}_{HD}$ & $\mathcal{L}_{Dice}$+ $\mathcal{L}_{Size}$ &$\mathcal{L}_{Dice}$ + $\mathcal{L}_{clDice}$  \\ \hline
				
				WMH  & 1.04 $\pm$ 0.14  &  \cellcolor{beaublue}{0.98 $\pm$  0.17} &
				\cellcolor{beaublue}{1.01 $\pm$ 0.23} & 
				\cellcolor{beaublue}{0.91 $\pm$ 0.22} & 
				\cellcolor{red!10}{2.14 $\pm$ 1.26} \\ 
				Isles  & 0.69 $\pm$ 0.19  &  \cellcolor{beaublue}{0.48 $\pm$  0.04} &
				\cellcolor{beaublue}{0.57 $\pm$ 0.19} & 
				\cellcolor{red!10}{1.34 $\pm$ 1.08} & 
				\cellcolor{beaublue}{0.39 $\pm$ 0.10} \\ 
				Atrium & 0.25 $\pm$ 0.01  & \cellcolor{red!10}{ 0.29 $\pm$  0.02} &
				\cellcolor{red!10}{0.32 $\pm$ 0.01 }& 
				\cellcolor{red!10}{0.28 $\pm$ 0.03} & 
				\cellcolor{red!10}{0.28 $\pm$ 0.03} \\ 
				Colon & 0.17 $\pm$ 0.02  &  \cellcolor{beaublue}{0.13 $\pm$  0.01} &
				\cellcolor{beaublue}{0.13 $\pm$ 0.01} & 
				\cellcolor{beaublue}{0.13 $\pm$ 0.01} & 
				\cellcolor{red!10}{0.18 $\pm$ 0.03} \\ 
				Spleen & 0.22 $\pm$ 0.05  & \cellcolor{red!10}{ 0.24 $\pm$  0.09} &
				\cellcolor{beaublue}{0.09 $\pm$ 0.01} & 
				\cellcolor{beaublue}{0.18 $\pm$ 0.13} & 
				\cellcolor{beaublue}{0.12 $\pm$ 0.15} \\ 
				\hline
				Hippocampus-\textit{H1}	   & 3.76 $\pm$ 0.14  &  \cellcolor{beaublue}{1.81 $\pm$  0.14} &
				\cellcolor{beaublue}{2.67 $\pm$ 0.11} & 
				\cellcolor{beaublue}{2.88 $\pm$ 0.39} & 
				\cellcolor{beaublue}{1.30 $\pm$ 0.96} \\ 
				Hippocampus-\textit{H2}  & 0.95 $\pm$ 0.01  &  \cellcolor{beaublue}{0.23 $\pm$  0.01} &
				\cellcolor{beaublue}{0.87 $\pm$ 0.10} & 
				\cellcolor{beaublue}{0.74 $\pm$ 0.06} & 
				\cellcolor{beaublue}{0.10 $\pm$ 0.03} \\  \hline
				Prostate-\textit{CG} & 8.96 $\pm$ 3.33  &  \cellcolor{red!10}{9.05 $\pm$  3.33} &
				\cellcolor{red!10}{8.89 $\pm$ 3.11} & 
				\cellcolor{red!10}{8.78 $\pm$ 2.79} & 
				\cellcolor{red!10}{8.98 $\pm$ 3.32} \\ 
				Prostate-\textit{PZ} & 0.36 $\pm$ 0.13  &  \cellcolor{beaublue}{0.23 $\pm$  0.09} &
				\cellcolor{red!10}{0.33 $\pm$ 0.07} & 
				\cellcolor{red!10}{0.80 $\pm$ 0.12} & 
				\cellcolor{beaublue}{0.26 $\pm$ 0.11} \\
				\hline\hline
				ACDC-\textit{RVC}  & 0.18 $\pm$ 0.03  &  \cellcolor{beaublue}{0.16 $\pm$  0.00} &
				\cellcolor{beaublue}{0.13 $\pm$ 0.02} & 
				\cellcolor{beaublue}{0.11 $\pm$ 0.04} & 
				\cellcolor{beaublue}{0.12 $\pm$ 0.03} \\ 
				ACDC-\textit{MYO} & 0.04 $\pm$ 0.02  &  \cellcolor{red!10}{0.06 $\pm$  0.01} &
				\cellcolor{red!10}{0.08 $\pm$ 0.03} & 
				\cellcolor{red!10}{0.07 $\pm$ 0.03} & 
				\cellcolor{red!10}{0.06 $\pm$ 0.02} \\ 
				ACDC-\textit{LVC} & 0.06 $\pm$ 0.01  &  \cellcolor{red!10}{0.06 $\pm$  0.01} &
				\cellcolor{red!10}{0.07 $\pm$ 0.01} & 
				\cellcolor{red!10}{0.07 $\pm$ 0.02} & 
				\cellcolor{red!10}{0.05 $\pm$ 0.02} \\ 
				\hline
		\end{tabular}}
	\end{center}
	\label{tab:Results-MAECC}
\end{table*}

\section{Results and Analysis}\label{benchmark-results}

In this section, we report results of the benchmark datasets relative to the losses under consideration based on the training strategy explained in section \ref{benchmark-expSetting}. The segmentation performances are compared via the 2 usual segmentation metrics: the Dice score\cite{Crum2006a} (DSC) presented in \tablename~\ref{tab:BenchmarkResults-DSC} , the Hausdorff distance metric\cite{Beauchemin1998} (HD) presented in \tablename~\ref{tab:Results-HD}. In addition, we have computed the mean absolute error on the number of instances (connected components) presented in \tablename~\ref{tab:Results-MAECC}.



\subsection{Added value of prior losses over the Dice loss  baseline}
From the performance tables, we realize that there is always at least one prior-based loss that is superior to the Dice baseline (denoted by cells with blue background in the tables). Thus, the exploitation of prior-based losses generally has enhanced segmentation performance in 10 out of 12 anatomical objects of the 8 datasets.  For example, the Hausdorff loss has registered best performances on brain lesion segmentation tasks (WMH, Isles) and single-organ segmentation datasets. On the other hand, the boundary loss registered performances close to the best case performance on lesion tasks (Isles, WMH). The clDice registers best performances in 1 out of 3 multi-organ segmentation datasets and the size loss got good results on a selection of datasets including WMH, Atrium and Colon.

A close look at the Dice baseline performance over the entire set of datasets (first column in the tables), one can observe that the Prostate is quite challenging since it has the lowest Dice baseline performance. On the contrary, the ACDC dataset is the easiest with the highest Dice accuracy, and the problem of cardiac structure segmentation is well known and has been argued to be almost solved as demonstrated in paper\cite{Bernard2018}. Intuitively, an easy dataset would already register good performance given the simple Dice baseline and one would expect the addition of prior-based losses to have no added value, other than adding to the complexity of the training and degrading system performance. Indeed, the results benchmarked on the ACDC dataset registers little to no added value on the  performance relative to the baseline. Alternatively, if the dataset is too complex such as the case of the Prostate (multi-label segmentation, large organ size imbalance, large number of connected components), customized prior-based losses may be needed to accommodate its characteristics: almost no gain is obtained from prior losses for the Prostate dataset.

\subsection{Low-level vs. High-level Prior-based Losses}

Both Hausdorff and Boundary losses register good performances on most datasets and over all segmentation tasks: brain lesions, single-organs and multi-organ segmentation. The Hausdorff loss has a superiority over the Boundary loss in some dataset cases (Spleen, Hippocampus). For example, for the spleen dataset, the Hausdorff loss has registered best case performance in both dice accuracy (added value of 14 \%) and reduced the Hausdorff distances by over 30 \% in comparison to the Boundary loss. The superiority of the Hausdorff loss over the Boundary loss is mainly due to the fact that the Hausdorff loss extracts distance maps from both predicted and ground-truth contours, and minimizes the error between the two maps accordingly, whereas the Boundary loss simply fine-tunes the probability distribution via the ground-truth distance maps. Based on this, one can say that since Hausdorff targets optimizing the distance map entity directly between predicted and ground-truth labels, it can guarantee a better mapping between predicted outputs and the ground-truth than the Boundary loss. Despite this significance, the Hausdorff loss is very computationally expensive since it consists in computing the predicted distance maps online while training, which directly affects training time. Hence, one may consider that the Boundary loss may represent a reasonable trade-off between good segmentation performance and computational cost. 

\begin{figure}[t!]
	\centering
	\includegraphics[width=11cm, height=5cm]{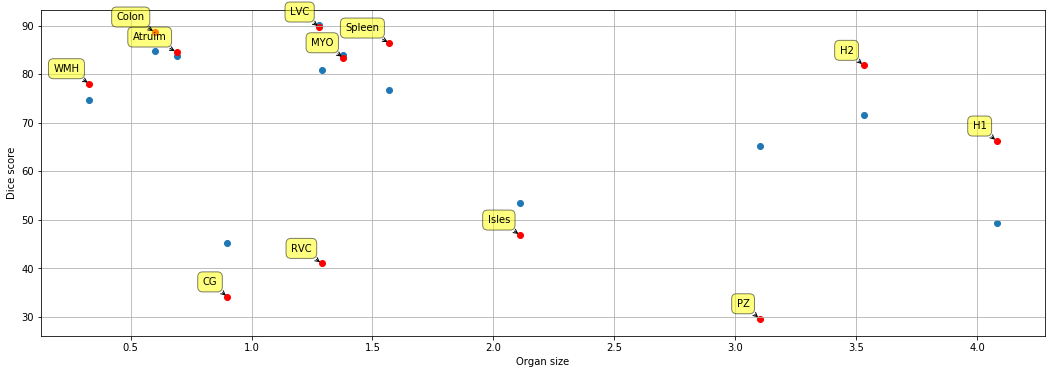} 
	\caption{Influence on the organ size on the average Dice score. Each dot represents a dataset. Blue (resp. red) dots show the Dice score obtained  with the Dice loss (resp. the Size loss).}
	\label{Size-Variation}
\end{figure}

Regarding the high-level prior-based losses, results are mixed: the size-based loss can either provide great improvement (e.g., WMH, Atrium, Colon), or much worse results (e.g., ISLES). For example, the size loss registers equivalent performance in the case of the WMH dataset relative to the best case segmentation result, but performs poorly on ISLES, despite the similarity in nature between the two datasets. We hypothesize that this may be due to the overall lesion sizes. A closer look at \tablename~\ref{tab:metadata} showing the meta-data characteristic, we can gather that, on small sized organs (e.g.: WMH, Atrium, Colon),  the size loss registers performance either better or equivalent to the Dice baseline. Given datasets that have large size variability (e.g., isles, Prostate, or ACDC), the exploitation of the size loss degrades segmentation performance. This is mainly due to the fact that, generally, the exploitation of the size loss allows the network to learn average sizes of the organs. In the same essence, based on the results, one can see that size loss can not accommodate multi-organ segmentations. The above observations are illustrated in \figurename~\ref{Size-Variation} showing the Dice performance relative to organ sizes. We note that the datasets where the size loss registered degraded results (red dots) are for those whose organ sizes are of large variability or that include multi-label segmentation.
Hence, despite the fact that the size loss was initially customized to accommodate weakly supervised segmentation, it may be useful in full supervision, when the anatomical objects under consideration are very small structures, and occupying a tiny percentage of the overall image as in the case of the WMH dataset. 

The clDice has a similar behavior but to a lesser extent. It generally registers better performance than the Dice baseline in most single-label segmentation cases and one multi-organ segmentation dataset. However, the clDice loss degraded performance on other datasets such as the WMH and the ACDC. Despite the equivalence in Dice accuracy between the Hausdorff loss and clDice loss on the Hippocampus dataset, the Hausdorff loss outperformed the clDice relative to the Hausdorff distance (Hausdorff loss is about 8 \% lower than clDice loss in Hausdorff distance). This indicates its ability to take into consideration shape and border specifications. The degraded performance of clDice on Hausdorff distance can be explained by the fact that the loss is based on the skeleton maps, which tends to blur boundary specifications for the sake of revealing topological properties. This limitation is further verified by the clDice with the Hausdorff distance results on the ACDC dataset. Thus, even when the clDice registered best ranked results relative to the Dice Accuracy, the Hausdorff distance is degraded, even lower than the Dice baseline with regards to the Myocardium, for instance. Given tasks with high border irregularities, such as lesions, failing to consider boundary specifications can hinder overall performance (e.g. case of brain lesion in the WMH dataset).


When studying other meta-data features such as the number of connected components, one can see that the exploitation of high-level prior-based losses does not have a great influence on the results (see \tablename~\ref{tab:Results-MAECC}). We hypothesize that this may be due to the fact that high-level prior-based losses are rather customized to serve a particular task, or satisfy a particular constraint. If the task at hand does not conform with the dataset characteristics or attributes, the prior-based loss may generally have no added value.


Overall, we can hypothesize that contour-based losses are  rather generic, and can be useful for enhancing segmentation performance on any type of dataset. However, if we are aiming at preserving a particular characteristic or anatomical property, a customized high-level prior-based loss may be a feasible solution. Thus, high-level losses may provide improvement; however, they are not very stable and can not be generalized to all datasets and tasks.



\subsection{Limitations of the Current Proposed Benchmark}

Despite our intuitive analysis with regards to some relationships between loss performance and dataset characteristics however, we admit to many limitations. For starters, the proposed benchmark can not be generic, as there are many existent prior-based losses that we fail to include: low-level prior \cite{Caliva2019, Yang2018a, Mosinska2018} , high-level topological \cite{Clough2019a} or shape prior \cite{Dolz2017a, Mirikharaji2018a} . Moreover, due to the fact that high-level prior-based losses are customized to target a particular property, providing means of comparison with respect to their effectiveness is subjected to debate. Another key component to take into consideration is their optimization algorithms. Many prior-based losses are discrete in nature; hence, they require particular optimization strategies in order to insure good performances. Our proposed benchmark is based on plugging the considered losses into a penalty-based Lagrangian optimization technique and training via stochastic gradient descent and the ADAM optimizer.  On the level of the datasets, despite some similarities between datasets (Lesion task: ISLES, WMH, task: Single vs. Multi), however, the datasets are rather very different, each given a set of characteristics and properties. Hence, there are a lot of variables to take into consideration, which makes the means of comparison often limited. Despite these limitations, presenting a benchmark that can test prior-based losses on different tasks and datasets is important, because it can give the reader an intuitive initial judgment on which loss to choose based on the considered requirements and datasets properties. 

\section{Conclusion}
In this paper, we proposed a benchmark of prior-based losses on medical image segmentation datasets. We provided intuitive explanations on a few existing relationships between prior-based loss significance and dataset characteristics. We summarized the paper's realizations as follows: the size loss is generally significant when considering datasets of small structures and limited size variability. The contour-based losses generally, and Hausdorff loss particularly,  accommodates objects of multi-structures and border irregularities. 

Future work includes expanding the proposed benchmark in order to encompass a broader perspective of losses. Moreover, we aim to add other metadata features, in order to better characterize the organ and the task at hand, develop robust similarity feature vectors between datasets for more accurate comparison and conduct meta-learning to predict loss ranks and outputs so as to  address the computational complexity issues between losses and their peers. 





\subsection*{Disclosures}
The authors have no relevant financial interests in this article and no potential conflicts of interest
to disclose.

\subsection* {Acknowledgments}
The authors would like to acknowledge the ANR (Project APi, grant ANR-18-CE23-0014) and the CRIANN  
for providing computational resources. This work is part of the WeSmile project funded by PHC VanGogh. 

\subsection* {Data, Materials, and Code Availability} 
The code used to conduct the  the benchmark is publically available on GitHub: \\
\href{https://github.com/rosanajurdi/Prior-based-Losses-for-Medical-Image-Segmentation}{https://github.com/rosanajurdi/Prior-based-Losses-for-Medical-Image-Segmentation}

%
\bibliographystyle{ieeetr}  
\bibliography{report}   


\listoffigures
\listoftables

\end{document}